\documentstyle[12pt]{article}
\begin{document}

\title{ Critical Dynamics in the Early Universe
\thanks{\it Invited talk given at the Journees Relativistes,
Amsterdam, May 1992. Conference Proceedings edited by W. A. van Leeuwen
to appear in Classical and Quantum Gravity}}

\author{ B. L. Hu \\
{\small Department of Physics, University of Maryland,
College Park, Maryland 20742, USA}}
\date{umdpp 93-57}
\maketitle
\begin{abstract}
Methods and concepts for the study of phase transitions mediated by
a time-dependent order-parameter field in curved spacetimes
are discussed. A practical
example is the derivation of an effective (quasi-)potential for the
description of `slow-roll' inflation in the early universe.
We first summarize our early results on viewing the symmetry behavior of
constant background fields in curved but static spacetimes as finite size
effect, and the use of derivative expansions for constructing effective actions
for slowly-varying background fields. We then introduce the notion of
dynamical finite size effect to explain how an exponential expansion of
the scale factor imparts a finite size to the system and how the symmetry
behavior in de Sitter space can be understood qualitatively in this light.
We reason why the
exponential inflation can be described equivalently by a scale transformation,
thus rendering this special class of dynamics as effectively static.
Finally we show how, in this view, one can treat the class of `slow-roll'
inflation as a dynamic perturbation off the effectively static class of
exponential inflation and understand it as a dynamical
critical phenomenon in cosmology.
\end{abstract}

\newpage

\section{Effective Action for Dynamic Order-Parameter Fields}
\par
In this talk I would like to report on some new thoughts on the question
of how to construct an effective action for a slowly-varying order-parameter
field for the description of a class of inflationary cosmologies \cite{Guth}
where the transition to the true vacuum takes place via a
gradually changing potential, such as the `slow-roll' type
in  new inflation \cite{newInf}. Knowledge of the exact form of the effective
action holds the key to a complete description of a phase transition.
One can deduce not only the qualitative features
(first or second order) but also the quantitative details
(mechanisms and processes).
There has been recent interest in understanding the nature and
construction of the effective
potentials for scalar fields in different cosmological spacetimes.
The interest is both theoretical, when viewed as a problem of finding
the infrared behavior
of quantum fields in curved spacetimes, and practical, when it is applied to
descriptions of phase transitions in the very early universe (from the Planck
time to the GUT time), or even the late universe (e.g.,
for electroweak phase transitions).\\

The work I am reporting to you now is a continuation of the program of study
I began almost ten years ago with Denjoe O'Connor and T. C. Shen on the
symmetry behavior of quantum fields in curved spacetime. The first stage
of our work centered on the simpler category of static spacetimes and constant
order-parameter fields. We found that both geometrical
and topological factors influence the infrared behaviour (IR) and often
can be thought of as finite size effects.
For a review of this first stage of our work, see \cite{HuMG4,HuOCfse,OCSteHu}.
The implication
of finite size effect on cosmological phase transitions have been discussed
for the  de Sitter and mixmaster universes in \cite{HuOCdeS,HuOCmix}.\\

Most theoretical studies of inflation are based on the effective
potential of the inflaton field in the de Sitter universe. It is not
difficult to carry out such a calculation, as the mode functions are known
explicitly \cite{IRdeS}. However,
one needs to be careful in correctly incorporating the influence
of the zero mode and the higher modes of the spectrum on the critical behavior
We used the  2-particle
irreducible (2PI) formalism \cite{CJT}, which involves the consistent
solution of an equation for the background field and one for the
two-point function\cite{HuOCfse}.
This helped to  decipher the
infrared behavior near the critical point (for a massless, minimally coupled
field in a symmetric vacuum). We actually carried out such a program for a
general class of product spaces, including cosmological spacetimes as well
as the Kaluza-Klein and the imaginary-time finite temperature field
theories. These techniques and results should be  useful for tackling
a wide range of problems not necessarrily related to curved spacetime.
A problem of current interest is to work out
the fine structure of the finite temperature
effective potential for Higgs fields near the symmetric vacuum
in the electroweak phase transition in connection with,  e.g., late-time
baryogenesis processes.\\

Although we can get a reasonable intuitive description of the
physics in the above situations, quantatative statements are still
unreliable because of the persistance of IR problems (even if the
2PI effective action is used). The essential reason for this is that
the microscopic renormalized parameters are no longer adequate for a
description of the IR physics. For example, in the finite temperature
case a perturbative analysis in terms of the $4-d$ parameters is
well known to be plagued with IR problems. These IR problems lead to
a breakdown of perturbation theory in terms of the $4-d$ parameters
and are symptomatic of the fact that one is trying to describe
essentially $3-d$ physics in terms of $4-d$ physics. In other words,
the effective degrees of freedom in the problem are changing as a
function of scale.
Since 1990 Denjoe O'Connor and Chris Stephens have developed a
quantative  formalism wherein these IR problems are
controlled. They have applied their techniques to a broad range of
problems in different areas of physics
with new findings on dimensional reduction and cross-over behaviors.
The essence of their work is the development of a
renormalization group (RG) that can interpolate between qualitatively
different degrees of freedom\cite{OCS1,OCS2}. For the finite
temperature case their RG  is explicitly temperature ($T$) dependent,
and acts in such a way that for $T$ near zero it effectively integrates out
$4-d$ degrees of freedom and at high temperatures $3-d$ degrees of
freedom. Such an RG follows as closely as possible
the action of the dilation generator of true scale changes.
(See their contributions in this volume for a recent summary.)\\

The second stage of our work on phase transitions in the early universe
with time-varying background fields began in 1987 with Sukanya Sinha and
Yuhong Zhang.
We concentrated on situations where the order parameter field changes
either with space or time. A familiar example in condensed matter physics
is anisotropic superconductivity where one can use a gradient expansion
in the Landau-Ginzberg-Wilson effective potential to account for the
differences
coming from the next-to-nearest neighbor interactions.
For cosmological problems,
it is the time-dependence of the background field which one needs to deal with.
Strictly speaking, phase transition studies usually carried out
assuming a constant field in the de Sitter universe
is unrealistic, in that it only addresses the situation after the universe
has entered the inflationary stage and inflates indefinitely. This model cannot
be used to answer questions raised concerning the likelihood that
the universe will still inflate
if it had started from a more general, less symmetric initial state,  such
as the mixmaster universe \cite{Wald,BGH,SheHuOC}. Nor can one use this
model to study the actual process of phase transition (e.g., slow roll-over),
and the problem of exit (graceful or not) to the
`true' Friedmann phase. To do this, as is well-known, one
usually assumes that the potential is not exactly flat, but has a downward
slope which enables the inflaton field
to gradually (so as to give sufficient inflation)
settle into a global ground state. The cosmological
solution is, of course, no longer
a de Sitter universe.  Thus for a realistic description
of many inflationary transitions one needs to treat the case of a nonflat
potential and a time-varying field. The form of the potential and the
metric of the background spacetime determine
the behavior of the scalar field in the Laplace-Beltrami equation, but the
field in turn provides the source of the Einstein equation which determines
the behavior of the background spacetime metric. Hence they ought to be solved
self-consistently. (One usually considers  only the homogeneous mode
of the scalar field for the dynamics of inflation and the
inhomogeneous modes of quantum fluctuations for structure
formation.)
At the classical level, the wave equation
for the background scalar field (assumed homogeneous)
with self-interaction potential $V(\phi )$ in a
spatially-flat Robertson-Walker (RW) spacetime is given by

\begin{equation}
\ddot \phi + 3H(t) \dot \phi + V'(\phi ) = 0
\end{equation}
$$
\dot H + 3 H^2 = 8 \pi G V (\phi), ~~~
\dot H = - 4 \pi G \dot \phi ^2
$$

\noindent where $H(t) \equiv { \dot a / a }$ is the Hubble rate,
a dot denotes derivative with respect to cosmic time $t$, and a prime
denotes a  derivative taken with respect to its argument.
A trivial but important solution to these equations is obtained by assuming
that
$V(\phi) = V_0 =constant$,
$ \phi =  \phi _0 =constant$ and $H=H_0=constant$,
which is  the de Sitter universe $a= e^{H_0t}$
with a constant field. A less trivial but useful solution is the so-called
`power-law' inflation models \cite{PowInf}, with an exponential potential
and a slowly-varying inflaton field [see Eqs.(3, 4) below]. One can
take this as an example of a `slow-roll' transition.
The methods we have devised can be used to derive the effective
potential (strictly speaking, quasi-potential) of such classes of spacetimes
and fields. Let me describe in the following sections
the theoretical difficulties and ways to overcome them. Records
of the second stage of our research can be found in \cite{dfse,SinHu,cgea}.
\footnote{
Before closing this Introduction, let me make a comment on the meaning of the
term `critical dynamics' as used in the context of cosmological
phase transitions. By it we
refer to studies of phase transitions mediated by a time-dependent
order parameter field in contrast to static critical phenomena
where the order-parameter field is  constant in time. We are using this term
in a general sense, not necessarily referring to the specific conditions of
critical phenomena as discussed in condensed matter systems \cite{CriDyn}.
For example,
critical phenomena usually deals with the change of the order parameter
field near the critical point as a function of temperature.
In cosmology, temperature $T$ is a parameter usually (e.g. under the assumption
of adiabatic expansion) tied in with the scale factor $a$ and does not play
the same role as in critical phenomena. In the new
inflationary scenario, the critical temperature $T_{c}$ is defined as the
temperature at which a global ground state (the true vacuum) first appears.
The stage
when vacuum energy begins to dominate and inflation starts is the commencement
of phase transition.  The stage when the system begins to enter the true vacuum
and reheat can be regarded for practical purpose as the end of phase
transition.
Throughout the process of inflation the system is in a `critical' state.
The progression of a cosmological phase transition is measured not by
temperature, but by
the change of field configurations in time. Criticality corresponds to
the physical condition that the correlation length
$\xi  = m^{-1}_{eff} \rightarrow  \infty $, or
$ m^2_{eff} = {d^2V / {d {\phi}^{2}}}|_{ \phi=0}\rightarrow  0$
which  may or may not be possible).
In critical dynamics studies of  condensed matter systems
one usually analyses the time-dependent Landau-Ginzberg equation, with a
noise term representing the effect of a thermal bath and studies how the
system (order-parameter field) settles into equilibrium as it approaches the
critical point. We are not concerned with the corresponding cosmological
problem here.
An attempt to describe this aspect of inflationary transition
was made in \cite{CorBru}. See also \cite{HPR}.}

\section{Quasistatic Approximation vs Dynamical Finite Size Effect}

The effective potential $V(\phi)$ gives a well defined description of
phase transition only for a constant background (order-parameter) field.
If the order-parameter field is dynamic, the effective potential is ill-
defined and a host of problems will arise. Indeed,
the very meaning of phase transition can become questionable.
This is because as the field changes the effective action functional changes,
and the locations of the minima change also. The notion of symmetry breaking
and
restoration is meaningful only when there exist well-defined global and local
minima which does not change much in the time scale of the phase transition.
Changing background field will also
engender particle creation, which affects
the nature and energetics of phase transition as well.
Therefore, in the context of phase transitions
involving dynamic fields, short of creating a new framework,
one can at best discuss the problem in a perturbative
sense, where the background field is nearly constant (quasi-static),
so that an effective quasi-potential can still be defined \cite{qEffPot,SinHu}.
An effective Lagrangian for a slowly-varying background field can be obtained
by carrying out a quasilocal expansion in derivative orders of the field,
the leading term being the effective potential \cite{HuOC84}.
\begin{equation}
\cal L =
\cal L (   \phi, \partial_ \mu   \phi, \partial_\mu \partial_\nu  \phi, ...)
\end{equation}

One can
use this method to derive effective quasi-potentials for scalar fields in
flat space (For an example of its application to electroweak finite temperature
transition see the recent paper by Moss et al \cite{MTW}), or (in conformal
time)
for the conformally-flat Robertson-Walker spacetimes.
This is useful for studying cosmological phase transitions where the background
spacetime changes only gradually, as in the Friedmann (low-power law) solutions
$a= t^p, p<1$. (For a description see \cite{SinHu}.)
However, for the inflationary universe where the scale factor
undergoes rapid expansion following
either an exponential $a=e^{Ht}$ or a high power-law behavior,
the quasilocal expansion which assumes that the background field varies
slowly is usually inadequate.
That was the quandary we were in
until the idea of using scaling to describe inflation dawned upon us.
Viewed in this new light the de Sitter exponential expansion can be regarded
as effectively static.

The first lead to such a connection came from our earlier
investigation into the infrared behavior of quantum fields in
de Sitter universe.
If one follows the main results we obtained for static spacetimes
 and view the de Sitter space
in the $S^4$ coordinatization, one can easily come up with a fairly good
qualitative
depiction. In our earlier work we introduced the concept of an
effective infrared dimension (EIRD) \cite{HuOCfse}. This concept
has been generalized and given a quantitative meaning in the work of
O'Connor and Stephens \cite{OCS3} where they define an effective
dimension ($d_{eff}$) which for de Sitter space is a function of ${\eta \equiv
\xi /a}$ (see footnote 1 for the definition of $\xi$)
and varies between $4$ and a number
close to zero. The reason it cannot go to zero is because there is a
maximum value for the correlation length in de Sitter space and the
RG must stop running at this value. This in fact is a generic
feature of the RG for totally finite systems.
However, if one views de Sitter space in other
coordinatizations, such as the $S^3 \times R^1$ or the $R^3 \times R^1$ cases,
one would have quite a different description of the physics where
the obvious connection with a finite geometry is lost.
We know that physics should be the same despite
differences in coordinate descriptions.
The resolution of this puzzle brings in an interesting point on the
effect of spacetime dynamics on the symmetry behavior of a quantum field.
Specifically, for the special case of exponential expansion,
in the spectrum of the 4-dim (spacetime) wave operator,
there is a gap which separates the zero mode from the rest. This is what
gives the   $d_{eff}\simeq 0$ for the deep IR behavior of
the scalar field in these other coordinate descriptions.
Physically this arises from the fact that at late times, as a result of
exponential expansion, most of the high-lying
modes are stacked-up against the zero mode by the rapid red-shifting.
(The $S^3 or R^3$ spatial sector also becomes inmaterial.) The
appearance of a scale (the event horizon $H^{-1}$)
is a unique feature  of this exponential class of expansion.
It gives rise to effects identical to that originating from some
finite size in some
associated static spacetimes. This is why we refer to these effects
as `dynamical finite size effects' \cite{dfse,OCSteHu}.

\section{Inflation as Scaling: Static Critical Phenomena}

The other lead came from the work I did with Yuhong Zhang in 1990 \cite{cgea}
on coarse-graining
and backreaction in stochastic inflation. There, in trying to compare the
inflationary universe with phase transitions in the Landau-Ginzberg model,
using a $\lambda \phi^4$ field as example, we realize that the exponential
expansion of the scale factor can be viewed as the system undergoing
a Kadanoff-Migdal scale transformation \cite{scaling}.
This can be seen as follows:
Consider a  spatially-flat RW metric with a constant scale factor.
This is just the Minkowski spacetime.  Let us consider an ordered
sequence of such static hyperspaces (foliation) with scales $a_{0}, a_{1},
a_{2}$,
etc parameterized by $t_{n} = t_{0} + n \Delta t, n = 0, 1, 2, \ldots $ .
These spacetimes have the same geometry and topology but differ only in the
physical scale in space.  One can always redefine the physical scale
length $x'_{(n)} = a_{n} x$ to render them equivalent.  If each copy has scale
length magnified by a fixed factor $s$ over the previous one in
the sequence, i.e. $a_{n+1}/ a_{n} \equiv s =e^{H\Delta t}$, we get exactly the
physical picture as in an eternal inflation. Here $s$ is independent of time.
After $n$-iterations i.e.
$a_{n}/a_{0} = e^{n(H\Delta t)}$, or, with a continuous parameter
$a(t) = a_{0}e^{Ht}$.  It is
important to recognize that $t$ can be any real parameter not necessarily
related to time.  In other words, time in this case plays the role of
a scaling parameter. It does not have to be viewed  as a dynamical parameter.
Thus for this special class of expansion, the dynamics of spacetime can be
replaced equivalently by a scaling transformation. In so doing one renders
eternal inflation into a static setting. By contrast, the larger class of
power-law expansion $a=t^\gamma$ does not possess this scaling property.
We see that $a_n/a_0= (1+ n \Delta t/t)^\gamma$ depends on time. Hence they
cannot be viewed as effectively static.
A useful parameter which marks the difference between these two classes of
dynamics is
$\zeta  = |\dot H|/ {H^2} = \ddot\alpha/\dot\alpha^2$,
where $\alpha \equiv ln a$, which can be regarded
as a `nonadiabaticity parameter' of dynamics:
the de Sitter exponential behavior with $\zeta= 0$ is `static',
the slow-roll with small $\zeta << 1$ is `adiabatic',
while the RW low-power-law with  $\zeta \approx 1 $ is `nonadiabatic'.
This rather unusual characteratization is nevertheless quite useful. It
captures
the essence of our considerations above in distinguishing between static vs
dynamical finite size effect and static vs dynamic scaling \cite{CriDyn}.
As distinct from the rather
unique de Sitter case, where only
one parameter, the scaling parameter $s$,
is needed for the description of the dynamics of spacetime and the field,
in the general class of RW dynamics,
two parameters are required : the  scaling parameter
$s$ which describes inflation, and a dynamic parameter $t$ which describes
the evolution of the field different from  the `static' (eternal
inflation) case.
These two parameters appear also in the
dynamical renormalization group theory description of dynamical critical
phenomena.

\section{`Slow-Roll' as Dynamical Critical Phenomena}

Using the conceptual framework introduced above, one can understand why
the particular subclass of high-power-law expansion
associated with an exponential potential can hence be viewed as quasi-static,
as it differs only slightly from the exponential expansion
in their qualitative behavior.
It is in this context that one can once again introduce the quasilocal
approximation to derive the effective action for scalar fields to depict this
more realistic `slow-roll' inflation, now carried out as
a quasilocal perturbation from the de Sitter space, which is viewed as
effectively static.

A classical solution  to (2) is given by  \cite{PowInf}:

\begin{equation}
V(\phi) = V_0 exp (-{\epsilon \over l_p} \phi), ~~~~~~
a(t)= a_0 (1+ H_0 \Delta t / \gamma)^\gamma
\end{equation}

\noindent Here $l_p = 1/ \sqrt{8 \pi G}$ is the Planck length,
$V_0= {{3 \gamma-1}
\over {\gamma}}{{H_0^2} \over {8 \pi G}}, \gamma= 2/\epsilon^2$ and
$\Delta t= t-t_0$. The subscript $0$ denotes the value at an initial time
$t_0$ where the de Sitter solution holds.
Note also that for $\gamma \rightarrow \infty$ or $\epsilon \rightarrow 0$, we
get the class of de Sitter solution $a(t) = a_0 e^{H_0(t-t_0)}$.
In that limit, the scaling parameter $s = {{a(t+t_0)} \over {a(t)}}
=(1+ {\sigma \over \gamma})^{\gamma} $ goes over to $e^{\sigma}$, where
$\sigma = H \Delta t$ .

The time-dependence of the scalar field $\phi$ associated with the above
solutions of $V$ and $a$ is given by \cite{PowInf}

\begin{equation}
\phi (t) = \phi _0 + \sqrt { \gamma /4 \pi G}~~ ln (1 + H \Delta t /\gamma)
\end{equation}

Now, following the rationale we suggested above, we can view this solution
as a `quasi-static' generalization of the de Sitter solution. Expanding
$\phi(t)$ for small time difference from $t_0$, we get,

\begin{equation}
\phi(t)-\phi_0 = ( \epsilon l_p H_0) \Delta t -({{\epsilon^3} \over {8}} l_p^2
H_0^2) (\Delta t)^2 + ...
\end{equation}

Here we can identify the coefficients of the first two
terms $(\Delta t)^n, n=1,2$ as the leading coefficients
in a quasi-local expansion of of the effective mass which involves
the background field \cite{HuOC84}. After such an identification,
it is easy to adopt the
well-established derivative expansion scheme for the calculation of the
quantum effective quasi-potential. The general result
is given in \cite{SinHu}.
There, as we recall, the effective Lagrangian contains the `kinetic energy'
terms as well as the radiative corrections arising from the varying
background field.  Details of the derivation and discussions on
the physical meaning
of these results in `slow roll-over' inflations are given in \cite {HuSin}.
(Kay Pirk has recently derived a quantum solution to (3) \cite{Pirk}.)

\section{Summary}

Let me summarize the main points brought up in this talk
by the following schematic diagram:

\vskip .5cm
\noindent Constant Field in Static or ~~~~~SCALING~~~~~Exponential Expansion
$a=
   e^{Ht}$\\
Conformally Static Spacetimes ----------------$>$   `Eternal Inflation' \\
(Finite Size Effect)~~~~~~~~~~~~~~~~~~~~~~~~~~~~~~~~~(Dynamical Finite Size
Effe
   ct)
   \\

{\it Quasilocal~~ Approximation~~~~~~~~~~~~~~ Derivative~~ Expansion}\\

\noindent Slowly Varying Field in~~~~~~~~~~~~~~~~~~~~~~~~High Power-Law
expansio
   n
   $a=t^p$\\
RW Universe (Low Power-Law) ----------------$>$   `Slow Roll-Over'\\

For slowly-varying background fields one can use the method of derivative
expansion to derive the quasilocal effective Lagrangian. Usually this makes
sense only for static (or conformally-static, like the RW) spacetimes.
However, one can view the special class of exponential expansion as
effectively `static'. This can be understood with  the ideas of `dynamical
finite size effect' and implemented by treating inflation as
`scaling' transformations. The `slow roll-over' type of phase transition
used in many inflationary models
can be viewed as a quasi-static case, and derived as a dynamic perturbation
from the de Sitter universe.
An example which this reasoning can be applied to
for the calculation of the effective `quasi-potential' is the case of
a high-power-law expansion with an exponential potential for the inflaton
field.
\\

\noindent {\bf Acknowledgements}\\
I would like to thank Profs. W. van Leeuwen and C. van Weert
for their hospitality
during this  well-organized, efficiently-run, multi-faceted and colorful
conference.
This report is based on work carried out jointly with
Sukanya Sinha and Yuhong Zhang with whom I have enjoyed many lively
discussions.
I would also like to thank
Denjoe O'Connor and Chris Stephens for sharing their important recent
findings on the renormalization group theory and its many interesting
applications.
Research in this work is supported in part by NSF under grant PHY91-19726.

\end{document}